\documentclass[twocolumn]{aastex62}

\usepackage{soul}
\graphicspath{{./}{figures/}}

\newcommand \sw{{\it Swift}}

\newcommand{\tc}{\theta_\mathrm{c}}
\newcommand{\tg}{\theta_\mathrm{\Gamma}}

\begin{document}
\title{Structured Jets and X-ray Plateaus in Gamma-Ray Burst Phenomena}
\correspondingauthor{Gor Oganesyan}
\email{gor.oganesyan@gssi.it}
\author{Gor Oganesyan}
\affiliation{Gran Sasso Science Institute, Viale F. Crispi 7,I-67100,L'Aquila (AQ), Italy}
\affiliation{INFN - Laboratori Nazionali del Gran Sasso, I-67100, L'Aquila (AQ), Italy}
\affiliation{INAF - Osservatorio Astronomico d'Abruzzo, Via M. Maggini snc, I-64100 Teramo, Italy}

\author{Stefano Ascenzi}
\affiliation{Gran Sasso Science Institute, Viale F. Crispi 7,I-67100,L'Aquila (AQ), Italy}
\affiliation{INAF, Observatory of Abruzzo, Via Mentore Maggini snc, 64100 Collurania, Teramo, Italy}

\author{Marica Branchesi}
\affiliation{Gran Sasso Science Institute, Viale F. Crispi 7,I-67100,L'Aquila (AQ), Italy}
\affiliation{INFN - Laboratori Nazionali del Gran Sasso, I-67100, L'Aquila (AQ), Italy}
\affiliation{INAF - Osservatorio Astronomico d'Abruzzo, Via M. Maggini snc, I-64100 Teramo, Italy}

\author{Om Sharan Salafia}
\affiliation{INAF - Osservatorio Astronomico di Brera, via Emilio Bianchi 46, I-23807 Merate (LC), Italy}

\author{Simone Dall'Osso}
\affiliation{Gran Sasso Science Institute, Viale F. Crispi 7,I-67100,L'Aquila (AQ), Italy}

\author{Giancarlo Ghirlanda}
\affiliation{INAF - Osservatorio Astronomico di Brera, via Emilio Bianchi 46, I-23807 Merate (LC), Italy}
\affiliation{INFN -- Milano Bicocca, Piazza della Scienza 3, I--20123, Milano, Italy}


\begin{abstract}
The first multi-messenger detection of a binary neutron star merger, GW170817, brought to the forefront the structured jet model as a way to explain multi-wavelength observations taken more than a year after the event. Here we show that the high-latitude emission from a structured jet can naturally produce an X-ray plateau in gamma-ray burst (GRB) light curves, independent of the radiation from an external shock. We calculate the radiation from a switched-off shell featuring an angular structure in both its relativistic bulk motion and intrinsic brightness. Our model is able to explain the shallow decay phase (plateau) often observed in GRB X-ray light curves. We discuss the possible contribution of the structured jet high-latitude emission to other distinctive features of GRB X-ray light curves, and its capability to explain the chromatic optical/X-ray light curve properties.
\end{abstract}

\section{Introduction} \label{sec:intro}
The follow-up of gamma-ray burts (GRBs) by the X-ray Telescope (XRT, 0.3-10\,keV) on board the {\it Neil Gehrels Swift Observatory} (\sw\ hereafter - \citet{Gehrels2004}) has revealed a rich morphology and diversity in their X-ray counterparts. However, systematic studies of the X-ray light curves have identified a canonical behaviour characterized by the presence of an early steep temporal decay, often followed by a shallow phase at nearly constant flux eventually turning into the characteristic temporal decay expected from an external shock \citep{Nousek2006,Zhang2006}.  

Typically, the steep decay phase ($F_{\nu} \propto t^{-\alpha_1}$ with $3 \le \alpha_1 \le 5$) lasts up to $\sim 10^{2}-10^{3}$ s \citep{Tagliaferri2005,Brien2006}. The shallow decay phase (or ``plateau", $F_{\nu} \propto t^{-\alpha_2}$ with $\alpha_2 \sim 0.5$ or shallower) can extend up to $\sim 10^4-10^5$ s, and it is present in a good fraction of GRBs \citep{Nousek2006,Zhang2006,Liang2007,Willingale2007}. Later, the evolution of the X-ray light curve transitions to a more ``canonical'' decay ($F_{\nu} \propto t^{-\alpha_3}$, with $\alpha_3 \sim 1-1.5$), sometimes showing a further steepening on timescales of a few days. While $\alpha_{3}$  is consistent with the temporal slope expected in the standard afterglow model \citep{Sari1998}, the steep decay and the plateau phases call for a different interpretation.

The initial steep decay phase observed by XRT is often modelled as high-latitude emission (e.g. \citealt{Liang2007b}), i.e. the radiation received from larger angles relative to the line of sight, when the prompt emission from a curved surface is switched off (\citealt{Fenimore1996}). It was shown that high-latitude emission from a spherical surface has a power-law decay $F_{\nu} \propto t^{-(\hat{\beta}+2)}$, where $\hat{\beta}$ is the slope of the emitted spectrum, typically modelled as a simple power law, $F_{\nu} \propto \nu^{-\hat{\beta}}$ \citep{Kumar2000}. Different modifications of the standard high-latitude emission have been considered in previous studies. They include the effects of a finite cooling time \citep{Qin2008}, a non-power law input spectrum \citep{Zhang2009}, time-dependent bulk motion \citep{Uhm2015}, a finite size of the emitting shell with multi-pulse contributions \citep{Genet2009}, an off-axis observer \citep{Lin2018}, and inhomogeneities of the relativistic jet  \citep{Dyks2005,Yamazaki2006,Takami2007}. While these studies were focused on the effects of jet structure, the attention was restricted to the steep decay phase.

The plateau phase instead is usually explained by introducing non-trivial modifications of the standard afterglow theory. The proposed solutions include scenarios with time-varying microphysical parameters of the external shock \citep{Ioka2006}, long-lived reverse shocks \citep{Uhm2007,Genet2007a}, delayed afterglow radiation from an in-homogeneous jet \citep{Toma2006,Eichler2006}, afterglow radiation in the thick-shell model \citep{Leventis2014}, baryon loading into the external shock from the massive outer shell \citep{Duffell2015}, delayed onset of the afterglow emission \citep{Kobayashi2007}, two-component jet model \citep{Jin2007}, photospheric emission from a long-lasting outflow \citep{Beniamini2017} or possible emission prior to the main burst \citep{Yamazaki2009}. Also, prompt emission scattered by dust grains \citep{Shao2007} and late-time prompt emission originated from less powerful shells with relatively smaller bulk Lorentz factors \citep{Ghisellini2007} were proposed  as alternative models.

However, the model with the largest consensus considers additional energy injection to the external decelerating shock \citep{Rees1998c,Zhang2006,Granot2006a,Nousek2006}. The additional energy would prevent the blast wave from decelerating, thus avoiding the typical afterglow decay, $F \propto t^{-1}$. The plateau phase, observed up to several $\times 10^4$ s requires long-lasting activity of the central engine, which can be provided either by the long-term evolution of the accretion disk around a black hole \citep{Kumar2008,Cannizzo2009,Linder2010} or by the spin-down power released by a newly-born millisecond spinning and highly magnetized neutron star \citep{Dai1998a,Dai1998b,Dai2004,ZhaMes2001, Yu2010,Metzger2011, DallOsso2011}. 

Despite many theoretical efforts to explain the origin of the X-ray plateau (with or without energy injection), there is not a clear consensus on its origin (see \citealt{Kumar2015} for a review). The main difficulty of these models is the lack of a robust explanation for the observed chromatic behaviour of X-ray and optical afterglow light curves (see \citealt{Fan2006}). 

The structured jet, {\it i.e.} a jet with an angular profile in both the bulk Lorentz factor and luminosity \citep{Lipunov2001,Dai2001,Rossi2002,Zhang2002}, has been invoked to explain the multi-wavelength observations of GRB170817A/GW170817 taken over one year (e.g \citealt{Ghirlanda2018}). Moreover, the structured jet has been shown to provide a natural explanation for the luminosity function of GRBs \citep{Pescalli2015}. In this work we test the idea that, along with the steep decay, the high-latitude emission during the prompt phase from a structured jet can produce the plateau observed in the lightcurve at late times. Qualitatively, the decreasing relativistic beaming of the emission at higher latitudes leads to a shallower light curve with respect to the uniform jet, for which the bulk Lorentz factor is constant throughout the emitting surface. Furthermore, the latitude-structured bulk motion results in an oblate geometry of the emitting surface which further extends the high-latitude emission in time. To be more specific, we seek for  long-lasting ($\sim 10^{2}-10^{4}$ s) flat segments in the light curves from high-latitude emission arising after $10^{2}-10^{3}$ s when the emitting source is switched off. 

\section{High-latitude emission from a structured jet}
We assume an expanding shell in vacuum. As a result of prior interaction with a surrounding medium (the envelope of the progenitor star or the ejecta cloud of the progenitor neutron star merger), different parts of the shell move with different velocities. We assume axisymmetry of the velocity relative to the azimuthal angle (in spherical coordinates) and that the observer is perfectly aligned with the center of the shell. 
At a given time in the rest frame, radiation is produced throughout the entire shell with an infinitesimally short duration. 

\begin{figure}[ht!]
\begin{center}
   {\centering
  \includegraphics[width=0.40\textwidth]{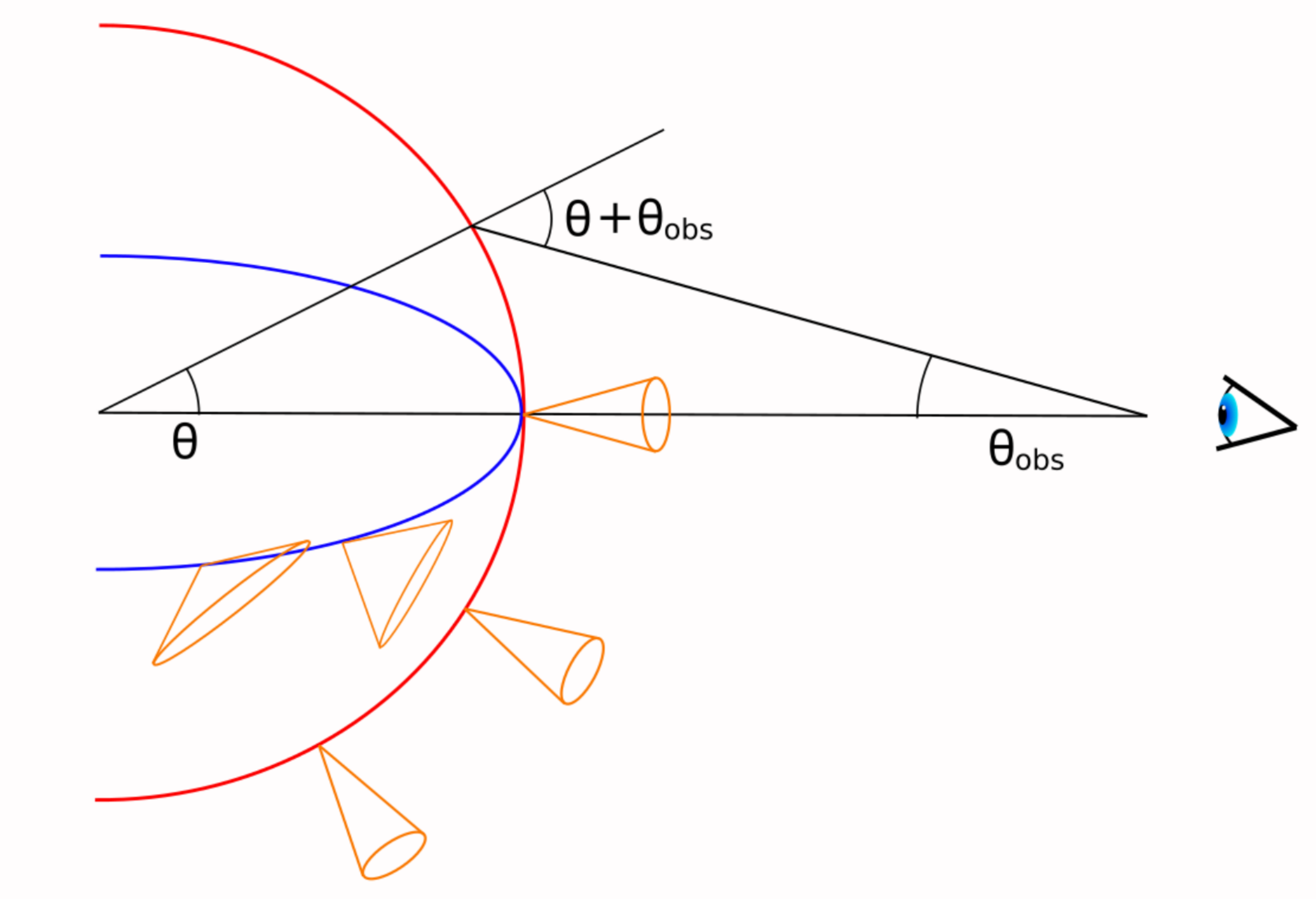}
  }
\caption{Spherical (red) and structured (blue) emitting surfaces corresponding to the uniform jet and the structured jet, respectively. For the uniform jet $\Gamma$ is constant all over the surface, while in the structured jet there is an angular dependence which results in the widening of the emitted radiation beaming cone. At a fixed $\theta$ the travel time of photons in the structured jet is longer than in the spherical case. We assume $\theta+\theta_{obs} \approx \theta$, i.e.~we consider distant observers. \label{fig:sketch}
}
\end{center}
\end{figure}

The adopted model is sketched in Fig.~\ref{fig:sketch}: we consider and compare the Uniform Jet model (red curve) with the Structured Jet model (blue curve). In the uniform jet model the Lorentz factor is constant along the emitting surface, which is a portion of a sphere. In the structured jet case, instead, the Lorentz factor $\Gamma(\theta)$ decreases with the angular distance from the jet axis, $\theta$.  \\

The observed flux then depends on the Doppler factor $\mathcal{D}(\theta)=1/[\Gamma(\theta)(1-\beta(\theta) \cos\theta)]$ where $\beta=v/c$. In the uniform jet model (red curve in Fig.~\ref{fig:sketch}), the bulk Lorentz factor is constant throughout the shell. This results in a beaming cone with the same angular size $\sim 1/\Gamma$ at each  $\theta$. 
With increasing time, the observer progressively receives further out of the beaming cone coming from annuli located at larger angular distances $\theta$, causing the monotonic decay of the received flux. In the structured jet model (blue curve in Fig.~\ref{fig:sketch}), the elongated geometry of the emitting surface causes a slower increase of $\theta$, which in turn increases the duration of the high-latitude emission. Moreover, the beaming cones corresponding to larger $\theta$ have increasingly wide beaming angles. 

With the assumption that each patch of the shell moves at a constant velocity, we derive the relation between the observed time $t_\mathrm{obs}$ and the angle of the jet $\theta(t_\mathrm{obs})$ from which radiation is received at $t_{\rm obs}$: 
\begin{equation}
\frac{1}{\beta(0)} - \frac{\beta(\theta)}{\beta(0)} \cos\theta = \frac{c t_\mathrm{obs}}{R_{0}}
 \label{eq:time}
\end{equation}
where $R_{0}$ is the radius of the emitting surface at $\theta=0$. The first photon is assumed to arrive from the origin of a GRB.

The observed specific flux is given by the intrinsic spectral shape $S(\nu')$ and the comoving time-integrated surface brightness $\epsilon(\theta)$: 
\begin{equation}
F_{\nu}(t_\mathrm{obs}) \propto  \mathcal{D}^{2}(\theta) \, S(\nu') \, \epsilon(\theta)  \, \frac{\beta^{2}(\theta)}{\Gamma(\theta)}  \, \frac{\sin\theta\cos\theta d\theta}{dt_{obs}} \, \vert_{\theta(t_\mathrm{obs})}
 \label{eq:flux}
\end{equation}
where the observer frequency is $\nu = \mathcal{D}(\theta)\nu'$ (primed quantities are in the comoving frame). The detailed derivation of Eq.~\ref{eq:flux} 
is given in Appendix~\ref{appendix}.

We assume that the intrinsic spectral shape $S(\nu')$ is angle-independent and we normalize it to 1. For both the uniform jet and the structured jet, the observed flux can be computed by Eq.~\ref{eq:flux},
once the bulk Lorentz factor $\Gamma(\theta)$, the comoving brightness   
$\epsilon(\theta)$ and the intrinsic spectral shape $S(\nu')$ are specified. For a uniform jet the brightness is constant throughout the surface.

It is worth noticing that the expression in Eq. \ref{eq:flux} integrates over equal arrival time rings, and thus represents a single dimension approximation for the computation of the flux. Considering a finite time duration of the pulse would require integration over equal arrival time surfaces (e.g.~see \citealt{Fenimore1996,Dermer2004,Genet2009,Salafia2016}). The latter results in second order contributions to the observed flux due to finite width of the emitting surface and it is neglected here.

\subsection{The Gaussian structured jet}
We adopt a Gaussian jet structure for $\Gamma(\theta)$ and $\epsilon(\theta)$, with the same form as in \citet{Salafia2015}: 
\begin{equation}
 \begin{array}{l}
  \epsilon(\theta) = \epsilon_c\;e^{-\left(\theta/\tc\right)^2}\\
  \Gamma(\theta) = 1 + (\Gamma_c-1)\;e^{-\left(\theta/\tg\right)^2}\\
 \end{array}
 \label{eq:Gaussian}
\end{equation}
where $\epsilon_c$ and $\Gamma_c$ are the comoving core brightness and the bulk Lorentz factor at $\theta=0$, while $\tc$ and $\tg$ are the jet scaling factors. \footnote{While the jet structure in \cite{Salafia2015} is adopted for the lab-frame energy per solid angle, we use it here for the comoving brightness.} To compute the observed flux density $F_{\nu}(t_{obs})$, we adopt a power-law spectral shape $S(\nu') \propto \nu'^{-\hat{\beta}}$. The shape of the spectrum influences the temporal behaviour of the high-latitude emission since the comoving energy range is blue-shifted by $\mathcal{D(\theta)}$, which introduces a spectral dependence in Eq.\ref{eq:flux}.
\begin{figure}[ht!]
\begin{center}
   {\centering
    \includegraphics[width=0.46\textwidth]{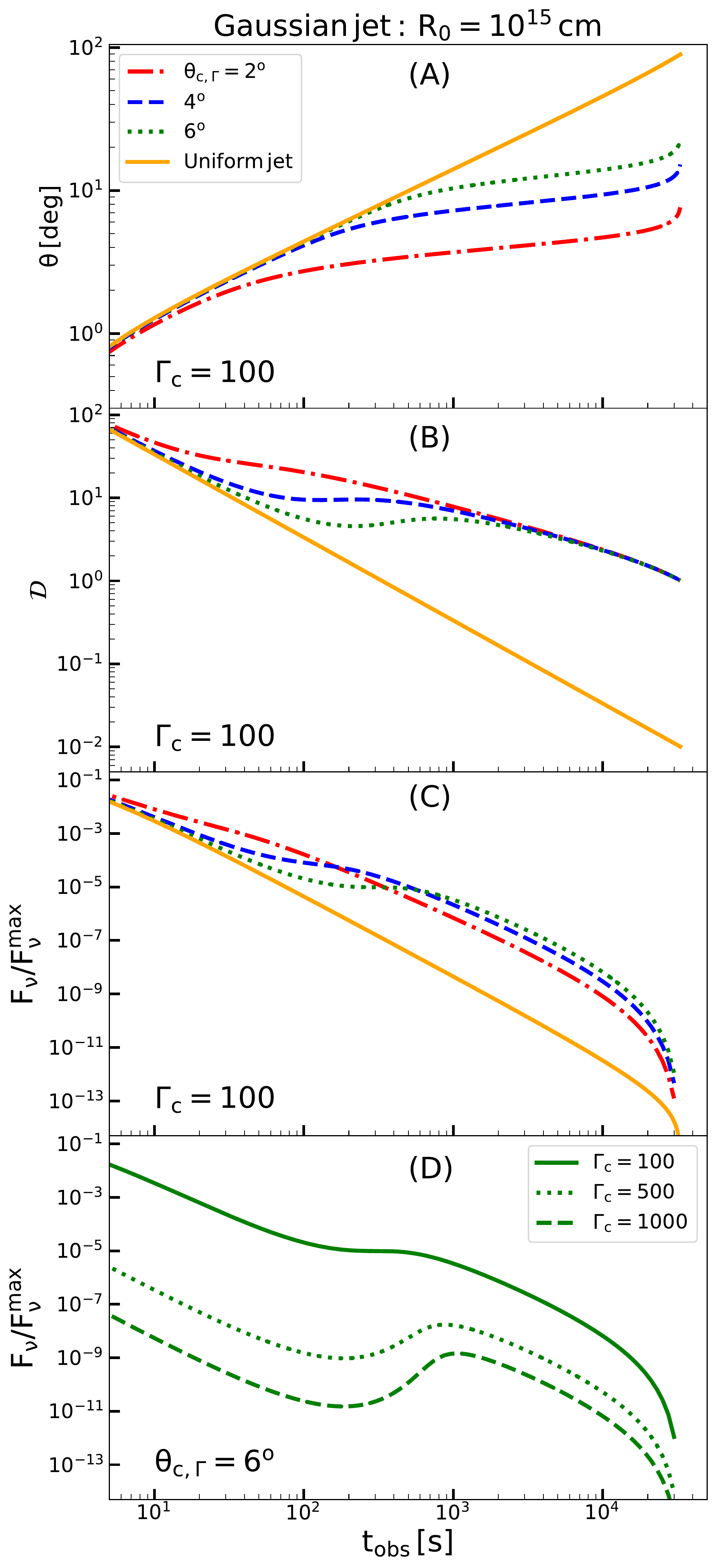}
  }
\caption{Time evolution of the polar angle $\theta$ of the jet from where radiation is observed (panel A), the corresponding Doppler factor $\mathcal{D}$ (panel B) and the high-latitude emission light curves ($\hat{\beta}=1$, panel C) for the Gaussian jet structure. Colors separate cases with different angular sizes of the jet for the same radial size $R_{0} = 10^{15}$ cm and central bulk Lorentz factor $\Gamma_{c}=100$. The solid orange lines represent the uniform jet. The dependence of the light curves from the central bulk Lorentz factor $\Gamma_{c}$ is shown in panel D ($\theta_{c,\Gamma}$ is fixed to $6^{o}$). \label{fig:Gaussian_Theta}}
\end{center}
\end{figure}

The shape of the light curve in the Gaussian jet structure case is defined by 5 parameters: $R_{0}$, $\Gamma_{c}$, $\theta_{c}$, $\theta_{\Gamma}$, and the spectral slope $\hat{\beta}$. We compute a sequence of light curves with 
variable parameter sets. For simplicity, we assume $\theta_{c}$ equal to $\theta_{\Gamma}$, and set $\hat{\beta}$ to 1.0,
which is the average observed spectral index in the plateau phase \citep{Liang2007}. We notice that flat portions of the light curves can be achieved at relatively late times $t_\mathrm{obs}>10^{2}$ s  for $R_{0} \ge 10^{15}$ cm, $\Gamma_c \ge 100$ and $\theta_{c,\Gamma} \ge 3^\circ$.


In Fig.~\ref{fig:Gaussian_Theta} we show the time evolution of $\theta$, $\mathcal{D}$ and the corresponding light-curves for fixed $R_{0} = 10^{15}$ cm and $\Gamma_{c} = 100$, and  three different values of $\theta_{c, \Gamma}$. The result in the uniform jet case is also reported for reference (solid orange line). The evolution of the polar angle $\theta$ from where the observer receives photons at a given time is shallower for the structured jet (panel A), as a consequence of the oblateness of the emitting surface: the narrower the jet, the longer the time travel difference between photons emitted at different angles. Therefore, the observed time-scale of high-latitude emission is longer than in the uniform jet case. The time evolution of the Doppler factor is affected both by $\theta(t_\mathrm{obs})$ and $\Gamma(\theta)$ (panel B). On one hand, the decrease of $\Gamma(\theta)$ with time increases $\mathcal{D}$ and on the other hand, the increase of $\theta$ with time reduces $\mathcal{D}$. As a result, there is a time-interval where $\mathcal{D}$ is roughly constant or slightly increasing, while $\epsilon(\theta)$ is still not decreasing much: this gives rise to the plateau in the light curve (panel C). The deviation of the Gaussian jet structure case from the uniform jet case happens essentially at the time when the observer sees the radiation coming from the core border, i.e.~when $\theta=\theta_{c,\Gamma}$. While we have fixed $\hat{\beta}$ to 1.0 and $R_{0}$ to $10^{15}$ cm, it is worth mentioning the effects of varying these parameters. Softer spectra result in steeper decays (as expected in the standard high-latitude emission) and fainter plateaus. The influence of the size of the jet on the high-latitude emission light curve is straightforward: smaller $R_{0}$ causes an earlier steep decay.

The dependence of the high-latitude emission on the bulk Lorentz factor (panel D) shows that the light curves with larger $\Gamma_{c}$ present a bump at $t_{\rm obs}>t_{c}$. This re-brightening can be interpreted as follows: given equal $\theta_{c, \Gamma}$, cases with higher $\Gamma_c$ are characterized by a higher beaming of radiation for $\theta < \theta_{c, \Gamma}$. Approaching $\theta_{c, \Gamma}$ a larger fraction of radiation will be beamed out from the observer, with respect to the low $\Gamma_c$ case. So, we can expect a higher flux when $\Gamma_c$ is lower. For $\theta \gg \theta_\Gamma$ instead, the steep decrease of $\Gamma(\theta)$, results in a widening of the beaming angle, which is similar for both the high and low $\Gamma_c$ cases. Therefore in this picture, the flux emitted by $\theta \gg \theta_\Gamma$ is almost the same for high and low $\Gamma_c$, while the flux emitted for $\theta \sim \theta_\Gamma$ will suffer a beaming-driven suppression in the high $\Gamma_c$ case. As a consequence, the lightcurve characterized by high $\Gamma_c$ will show the re-brightening, which is however the consequence of the suppression of the flux at earlier times.

The comoving brightness structure of the jet, $\epsilon(\theta)$, additionally suppresses the light curve intensity. The dependence of the light curve shape on $\epsilon(\theta)$ is much weaker than the dependence on $\mathcal{D}$. However, once the Doppler factor drops below 1, the emission becomes de-beamed and the plateau ends. The later very sharp drop is caused by having reached the edge of the jet. 

\subsection{Power-law structured jet}
\begin{figure}
\begin{center}
   {\centering
   \includegraphics[width=0.46\textwidth]{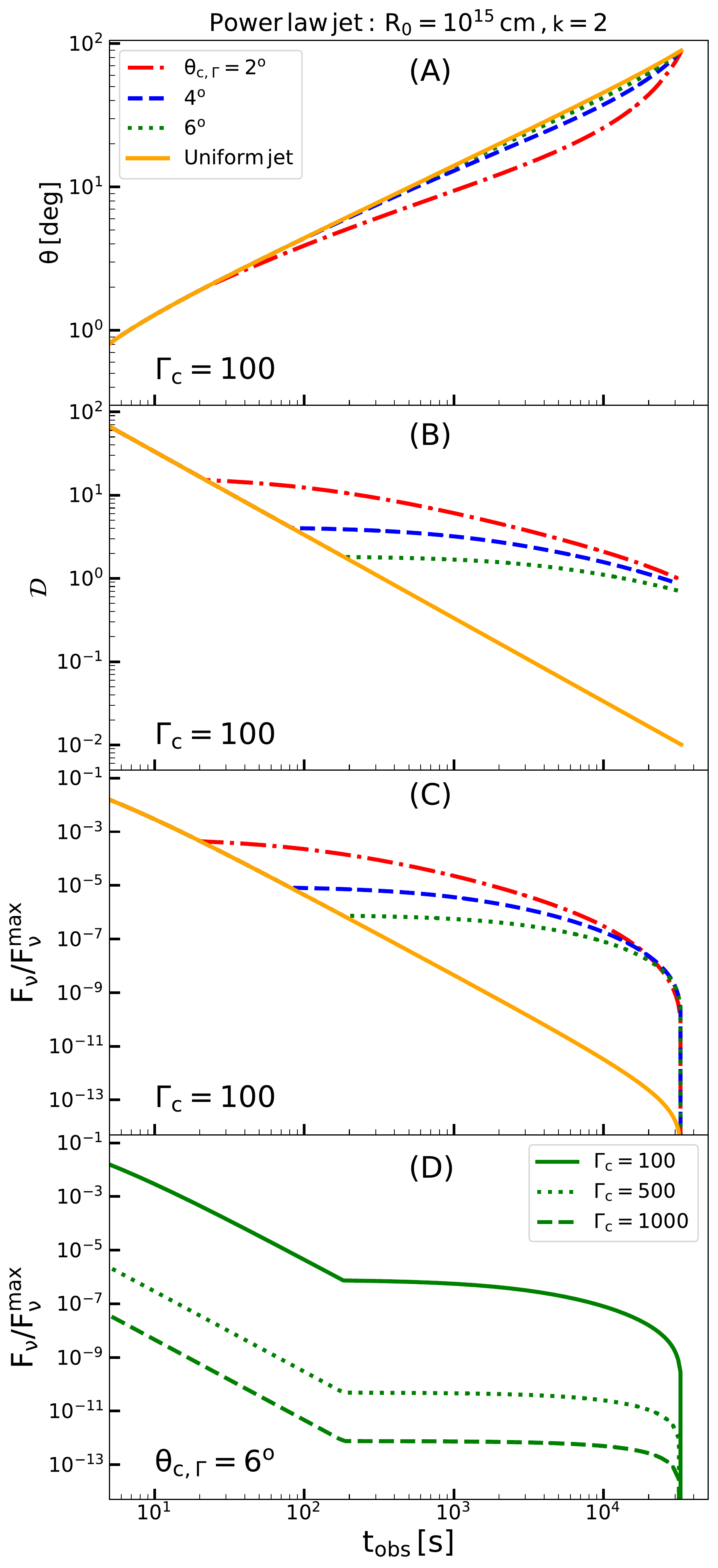}
  }
\caption{Time evolution of the polar angle $\theta$ of the jet from where radiation is observed (panel A), the corresponding Doppler factor $\mathcal{D}$ (panel B) and the high-latitude emission light curve ($\hat{\beta}=1$, panel C) for the power-law structured jet. Colors show cases with different angular sizes of the jet for the same radial size $R_{0} = 10^{15}$ cm, central bulk Lorentz factor $\Gamma_{0}=100$ and power-law index k=2. The solid orange lines represent the uniform jet. The dependence of the light curves from the central bulk Lorentz factor $\Gamma_{c}$ is shown in panel D ($\theta_{c,\Gamma}$ is fixed to $6^{o}$). \label{fig:Pl_Theta}}
\end{center}
\end{figure}
We compute high-latitude emission from the power-law jet structure for $\epsilon(\theta)$ and $\Gamma(\theta)$ adopted from \citet{Salafia2015}: 
\begin{equation}
 \begin{array}{ll}
\epsilon ( \theta ) = \left\{ \begin{array} { l l } { \epsilon _ { c } } & { \theta \leqslant \theta _ { \mathrm { c } } } \\ { \epsilon _ { c } \left( \theta / \theta _ { \mathrm { c } } \right) ^ { - k } } & { \theta > \theta _ { \mathrm { c } } } \end{array} \right.\\ \\ 
\Gamma ( \theta ) = \left\{ \begin{array} { l l } { \Gamma _ { c } } & { \theta \leqslant \theta _ { \Gamma } } \\ { 1 + \left( \Gamma _ { c } - 1 \right) \left( \theta / \theta _ { \Gamma } \right) ^ { - k } } & { \theta > \theta _ { \Gamma } } \end{array} \right. \\
 \end{array}
 \label{eq:PL}
\end{equation}
where $\epsilon_c$ and $\Gamma_c$ are the comoving brightness and the bulk Lorentz factor at $\theta=0$, $\tc$ and $\tg$ are the jet scaling factors and $\mathrm{k}$ is the slope of the power-law tail, i.e.~the steepness of the jet structure. 

We compute the observed flux density $F_{\nu}(t_{obs})$ using a power-law spectrum $S(\nu') \propto \nu'^{-\hat{\beta}}$ with $\hat{\beta}=1.0$. In the Fig.~\ref{fig:Pl_Theta} we show the time evolution of $\theta$ (panel A), $\mathcal{D}$ (panel B) and the corresponding light-curves for $R_{0} = 10^{15}$ cm, $\Gamma_{c} = 100$, $\mathrm{k}=2$ and $\theta_{c,\Gamma}$ varying in the range between $\rm 2^\circ$ and $\rm 6^\circ$ (panel C). We present also high-latitude emission from the uniform jet (solid orange line). One can notice that the rise of $\theta$ is faster compared to the Gaussian jet structure case. This is simply due to the shallower angle dependence of $\Gamma(\theta) \propto \theta^{-2}$ than in Gaussian jet structure. The Doppler factor deviates from the uniform jet case at times $t_\mathrm{obs}>t_{c}$ (post-core): the smaller the core size of the jet, the earlier is the slowing down of the $\mathcal{D}(t_{obs})$ variation. The major difference of $\mathcal{D}(t_{obs})$ compared to the Gaussian jet structure case is that the post-core temporal decay is monotonic, contrary to the ``bumpy'' trend of the Gaussian jet structure (see panel B of Fig.~\ref{fig:Gaussian_Theta}). The plateau phase is more prominent and longer (up to few $10^{4}$ s) in the the power-law jet structure case. The fast rise of $\theta(t_{obs})$ results in the sharp decay phase at the end of the plateau. 

We show the high-latitude emission dependence from $\Gamma_{c}$ in panel D. One can notice that the plateau phase is flatter with an increase of $\Gamma_{c}$. The flat and long plateau is provided by the shallow jet structure i.e. $k=2$ which causes an extremely fast rise of $\theta(t_{obs})$. 

\begin{figure}
    \centering
    \includegraphics[width=\columnwidth]{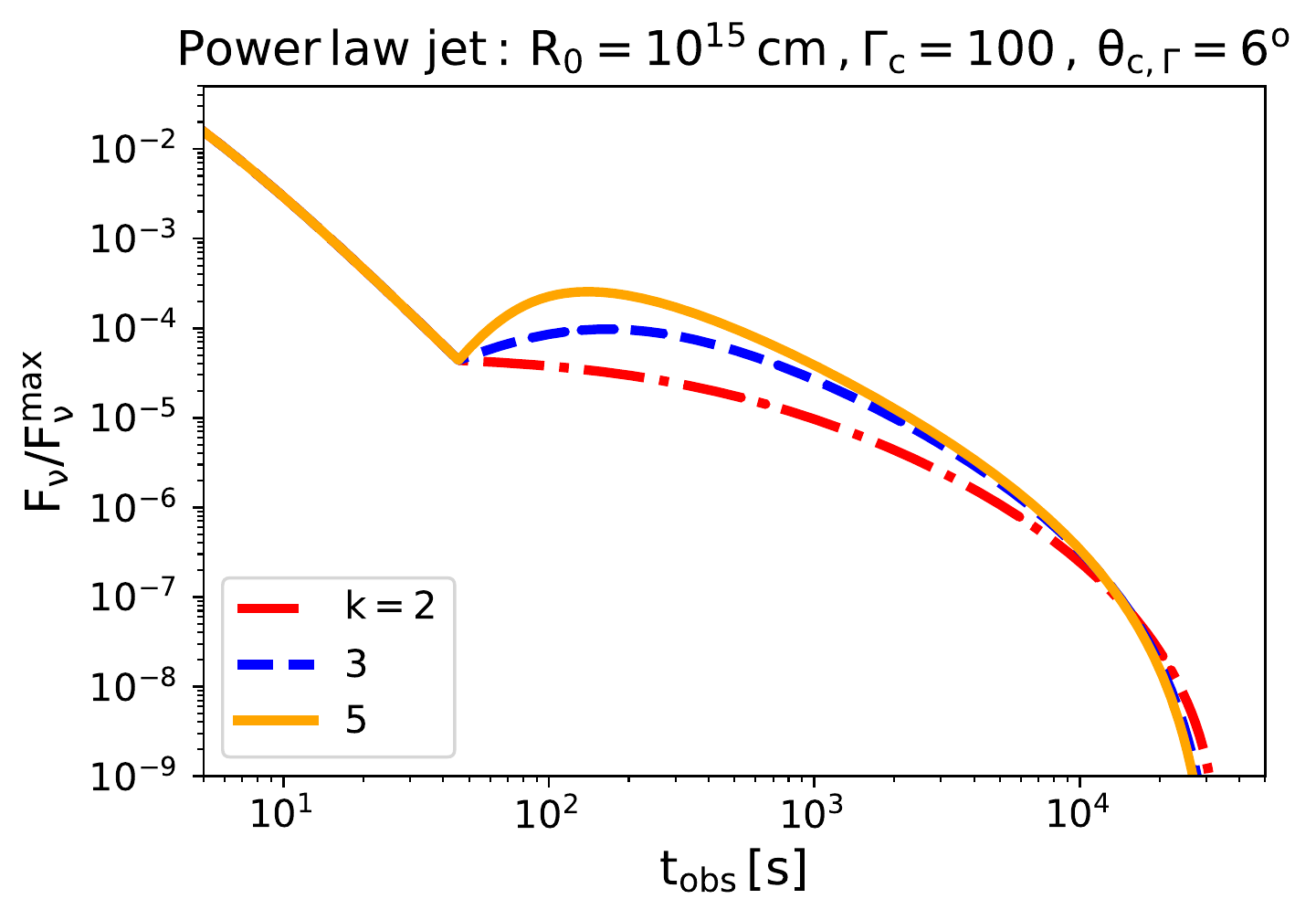}
    \caption{The high-latitude emission light curves for the power-law structured jet. Colors show cases with different power-law indices k.}
    \label{fig:k}
\end{figure}

The increase of k results in re-brightening of the light curve (see Fig.~\ref{fig:k}). 
The origin of the re-brightening for higher k is due to the fact that a steeper jet structure leads to a faster widening of the beaming angle. This results in a higher fraction of radiation reaching the observer at post-core time.

\section{Testing predictions of the high-latitude emission model}
In order to test our model ability to match  the temporal properties of the observed X-ray plateaus, we compare our model predictions with the observed X-ray light curves.
\begin{figure*}
\begin{center}
\includegraphics[width=0.45\textwidth]{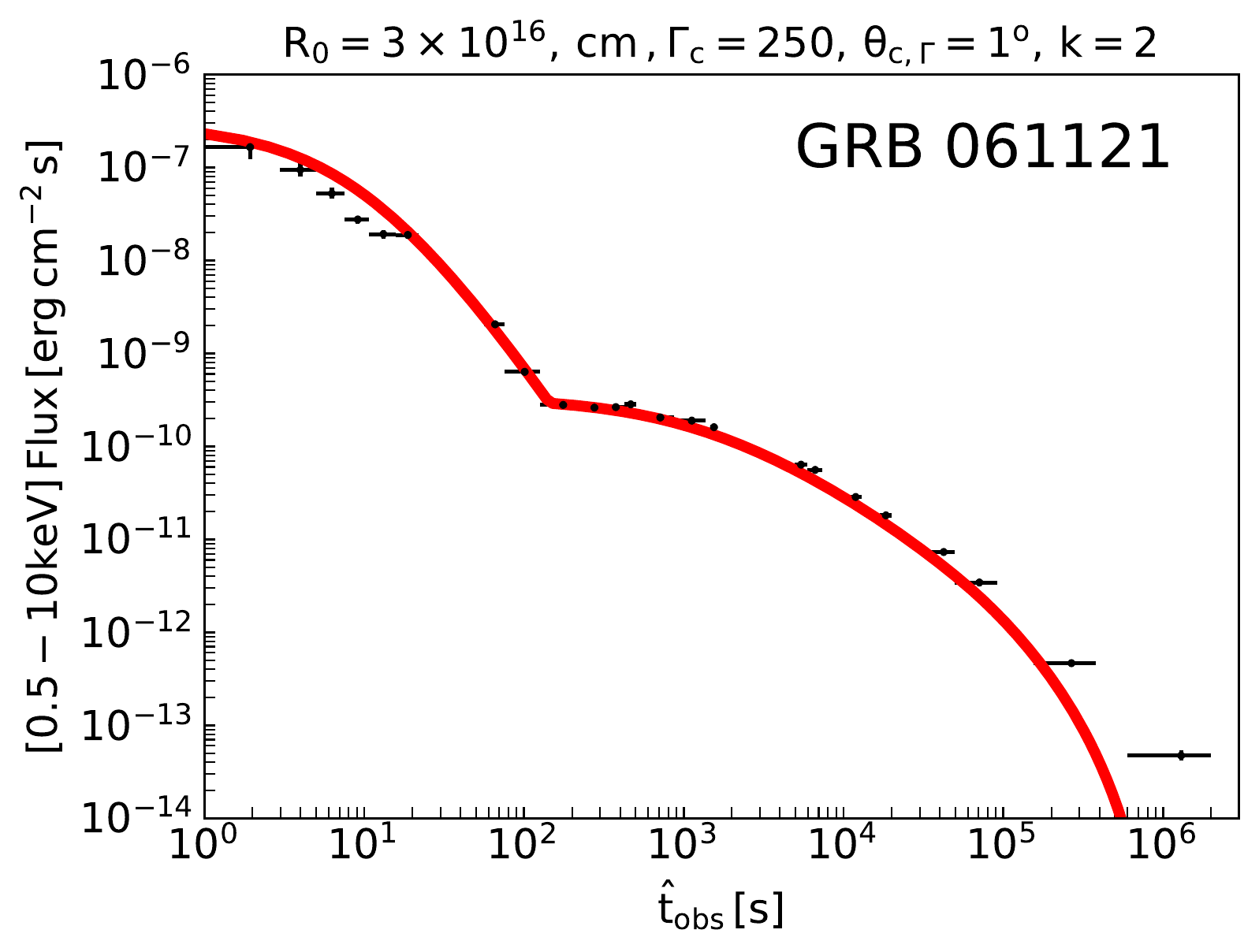}
\includegraphics[width=0.45\textwidth]{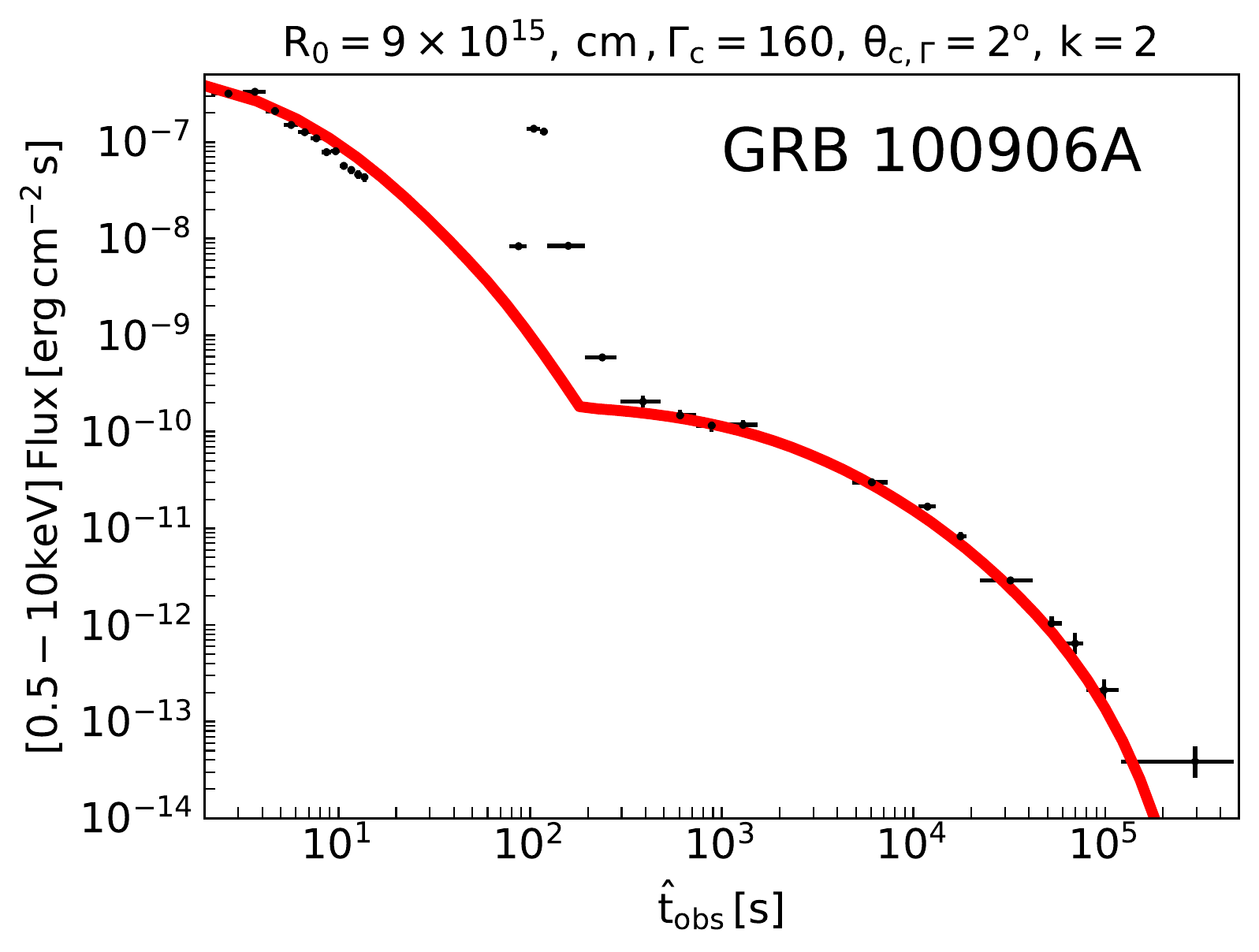}

\caption{X-ray light curves of GRB 061121 (left panel) and GRB 100906A (right panel) modeled by high-latitude emission with the power-law jet structure (red). To model the X-ray lightcurve of GRB 061121 we have adopted a power-law jet structure with $\Gamma_{c} = 250$, $R_{0} = 3 \times 10^{16} \rm cm$, $\theta_{c} = \theta_{\Gamma} = 1^\circ$, and $\mathrm{k}=2$. For 100906A we used $\Gamma_{c} = 160$, $R_{0} = 9 \times 10^{15} \rm cm$, $\theta_{c} = \theta_{\Gamma} = 2^\circ$, and $\mathrm{k}=2$. To accurately account for the steep decay phase, we have implemented the synchrotron-like spectra with the characteristic energies derived from the spectral fits around the peak time preceding the steep decay phase: for GRB 061121 we fixed the $E_{\rm break}$ to $\rm 3$ keV  and $E_{\rm peak}$ $> 150$ keV, while for GRB 100906A $E_{\rm peak}$ to $\rm 100$ keV and $E_{\rm break}$ to $\rm 0.1$ keV. \label{fig:061121}}
\end{center}
\end{figure*}

We compare our model calculations with the observed X-ray light curves of GRB 061121 \citep{061121_swift,061121_konus} and GRB 100906A \citep{100906_swift,100906_konus}. The choice of these particular GRBs is motivated by the following requirements: 
\begin{itemize}
    \item Presence of the fast decay and plateau phases in the X-ray light curve. 
    \item Relatively bright GRB in order to best meet the assumption of an ``on-axis" observer.
\end{itemize}

\subsection{High latitude emission against X-ray data}

For each the time-bin we downloaded XRT spectra from the UK Swift Science Data Centre at the University of Leicester\footnote{\url{https://www.swift.ac.uk/xrt_spectra/}}. We then fitted each of time-resolved XRT spectra using XSPEC  (v12.10.0c) by {\it a power-law} model taking into account Galactic and intrinsic metal absorption \citep{Wilms}. The Galactic absorption has been estimated from \cite{Kalberla} while the intrinsic column density of hydrogen is a free parameter of the fit. We have derived the unabsorbed flux for all time-resolved spectra and then built the final lightcurves of GRB 061121 and GRB 100906A shown in Figure~\ref{fig:061121}. The peak time of the main prompt pulse for GRB 061121 and GRB 100906A is at $t_0 \sim 74$ s and $t_0 \sim 7$ s, respectively. This is chosen as time zero for our model, {\it i.e.} we assume the emitting source is switched off at $t_{0}$. Therefore, we changed the reference times for the light curves correspondingly. For the initial 10 s of GRB 100906A which lacks the XRT data, we estimated the soft 
X-ray flux by extrapolating the Swift/BAT spectra to the XRT energy range\footnote{\url{https://www.swift.ac.uk/burst_analyser/}}.

In order to produce the high-latitude emission light curve, we assume a synchrotron spectrum of the prompt emission pulse. Synchrotron has been proposed as the dominant radiation mechanism responsible for GRB prompt emission \citep{Rees1994, Katz1994, Sari1997,Kobayashi1997,Daigne1998}. In a few GRBs prompt spectra were successfully fitted assuming a single-component synchrotron emission from a  non-thermal population of electrons (e.g. \citet{Tavani1996,Lloyd00,Zhang2016,Zhang2018}). Recent studies of the broad-band GRB spectra have shown the ability of the synchrotron model {\it alone} to account for the entire prompt emission spectrum once {\it the cooling of electrons is taken into account} \citep{gor2017,gor2018,gor2019, Mery2018,Mery2019}. 
Motivated by these recent results, we adopt a double broken power-law model (2BPL) with synchrotron slopes to model high-latitude emission from the prompt emission pulse. The photon indices below and above the low-energy break at $E_\mathrm{break}$ are $\alpha_1 \rm = -2/3$ and $\alpha_2 \rm = -1.5$ which correspond to spectral indices $\beta_1 \rm = -1/3$ and $\beta_2 \rm = 0.5$ in the $F_{\nu} \propto \nu^{-\hat{\beta}}$ representation.

The spectrum around the peak time of the prompt pulse ($t_{0}$) of GRB 061121 is best fitted with $E_{\rm break} \sim \rm 3$ keV \citep{gor2017}. We use this value as input for modelling the high-latitude emission in the XRT spectral range. The corresponding break energy in the comoving frame is then $E_{\rm break}/\mathcal{D}(\theta=0)$. Furthermore, we assume the peak energy (corresponding to the break at higher energy in the 2BPL model) $E_{\rm peak}>$ 150 keV since it is not constrained by the BAT data. However, we verified that its exact value does not affect the X-ray light curve at the observed time of the plateau. To reproduce the fast decay and the plateau phase observed in the X-ray emission (at 0.5-10 keV) of GRB 061121 using the high-latitude emission model, we adopt a power-law jet structure with $\Gamma_{c} = 250$, $R_{0} = 3 \times 10^{16} \rm cm$, $\theta_{c} = \theta_{\Gamma} = 1^\circ$, and $\mathrm{k}=2$ (red line in the left panel of Fig. \ref{fig:061121}). The final model is represented in the reference time $\hat{t}_{obs}$ of the first photon arriving from the head of the jet, {i.e.} $\theta=0$ (see Appendix~\ref{appendix}).

The high latitude model for GRB 100906A is obtained with $\Gamma_{c} = 160$, $R_{0} = 9 \times 10^{15} \rm cm$, $\theta_{c} = \theta_{\Gamma} = 2^\circ$, and $\mathrm{k}=2$ (the right panel of Fig.~\ref{fig:061121}). We adopted the peak energy of BAT spectrum $\sim$ 100 keV \citep{100906_auto}, placing the break energy at sub-keV energy range. One can notice an X-ray flare at $\sim$ 120 s during which the high latitude emission is well below the X-ray flux.

These examples illustrate the role of high-latitude emission in shaping GRB lightcurves: this additional component of the prompt emission results to be particularly important in the X-ray range to naturally produce a steep decay followed by a longer-lasting plateau as a consequence of a generic jet structure. The high-latitude emission model requires large radii of the prompt emission zone $\sim 10^{16}$ cm. Such radii are expected in the synchrotron-dominated models for the prompt emission \citep{Kumar2008a,Beniamini2013,Beniamini2018}.  Additional constrains on the jet properties would also require a comprehensive consideration of the observed GRB luminosity functions (see \citealt{Beniamini2019}).

There are observational indications from the temporal and spectral behaviour of multiwavelength data that optical and X-ray emission could arise from 
different emission regions (e.g. \citealt{Li2015}). This is addressed below.

\subsection{Multiwavelenght modelling including the forward shock contribution}

Here we investigate how accurately our model predictions can describe the multiwavelength emission of the GRB 061121 and GRB 100906A including the forward shock emission expected from the structured jet. We assume the model of \citet{Salafia2019}, where the afterglow emission from the forward shock accounts for the jet structure. The model based on standard afterglow concepts \citep{Meszaros1993,Meszaros1997,Sari1998}, accounts for synchrotron emission from the external forward shock, and includes the effect of radiative cooling and of synchrotron self-absorption. The shock dynamics is computed throughout its evolution, from coasting to the self-similar phase \citep{Blandford1976}, down to the non-relativistic phase. The emission at a given observed time is computed on the relevant equal-arrival-time surface (EATS), assuming the emitting region (i.e.~the shocked ISM material) to be geometrically thin. While the model itself can account for off-axis viewing angles, we assume here the observer to be on-axis, consistently with the assumption on the prompt emission phase, which speeds up the computation (since in that case we can exploit the azimuthal symmetry of the afterglow image and thus effectively reduce the EATS dimensionality by one). In order to compute the afterglow emission, we need to specify the jet kinetic energy profile $dE_\mathrm{K}/d\Omega(\theta)$. The simplest (and most widely used) assumption, which we adopt here, is to set the energy radiated in the prompt phase from each solid angle element to a constant fraction of the kinetic energy of the jet material\footnote{While being simple, this assumption actually conflicts with what one would expect, e.g., in the internal shock scenario: slower parts of the jet are presumably less efficient in converting kinetic energy into prompt emission. Accounting for this kind of effect would therefore go in the direction of making the decay in Eq.~\ref{eq:afterglow_structure} shallower.}. In our case this leads \citep[see Appendix A of][]{Salafia2015} to $dE_\mathrm{K}/d\Omega(\theta)\propto \beta^2(\theta)\epsilon(\theta)$. Thus, the angular dependence of the kinetic energy is approximately the same as that of $\epsilon$, namely
\begin{equation}
    \frac{dE_\mathrm{K}}{d\Omega} = \left\lbrace\begin{array}{lr}
        (E_\mathrm{c}/4\pi) & \theta\leq\theta_\mathrm{c}  \\
        (E_\mathrm{c}/4\pi)(\theta/\theta_\mathrm{c})^{-k} & \theta>\theta_\mathrm{c}
    \end{array}\right.\label{eq:afterglow_structure}
\end{equation}
in the power-law case, or
\begin{equation}
    \frac{dE_\mathrm{K}}{d\Omega} = \frac{E_\mathrm{c}}{4\pi} \exp\left[-\left(\frac{\theta}{\theta_\mathrm{c}}\right)^2\right]
    \label{eq:afterglow_structure_Gaussian}
\end{equation}
in the Gaussian case. Here $E_\mathrm{c}$ is the jet core isotropic-equivalent kinetic energy. The initial Lorentz factor profile is still given by Eq.~\ref{eq:PL}. The remaining relevant parameters for the afterglow phase are the external interstellar medium (ISM) number density $n$, the shock-accelerated electron power-law index $p$, the post-shock internal energy density fraction shared by the accelerated electrons $\epsilon_\mathrm{e}$ and the fraction shared by the magnetic field $\epsilon_\mathrm{B}$, all of which we assume to be independent from the angle.

\subsubsection{GRB 061121}

\begin{figure}
    \centering
    \includegraphics[width=\columnwidth]{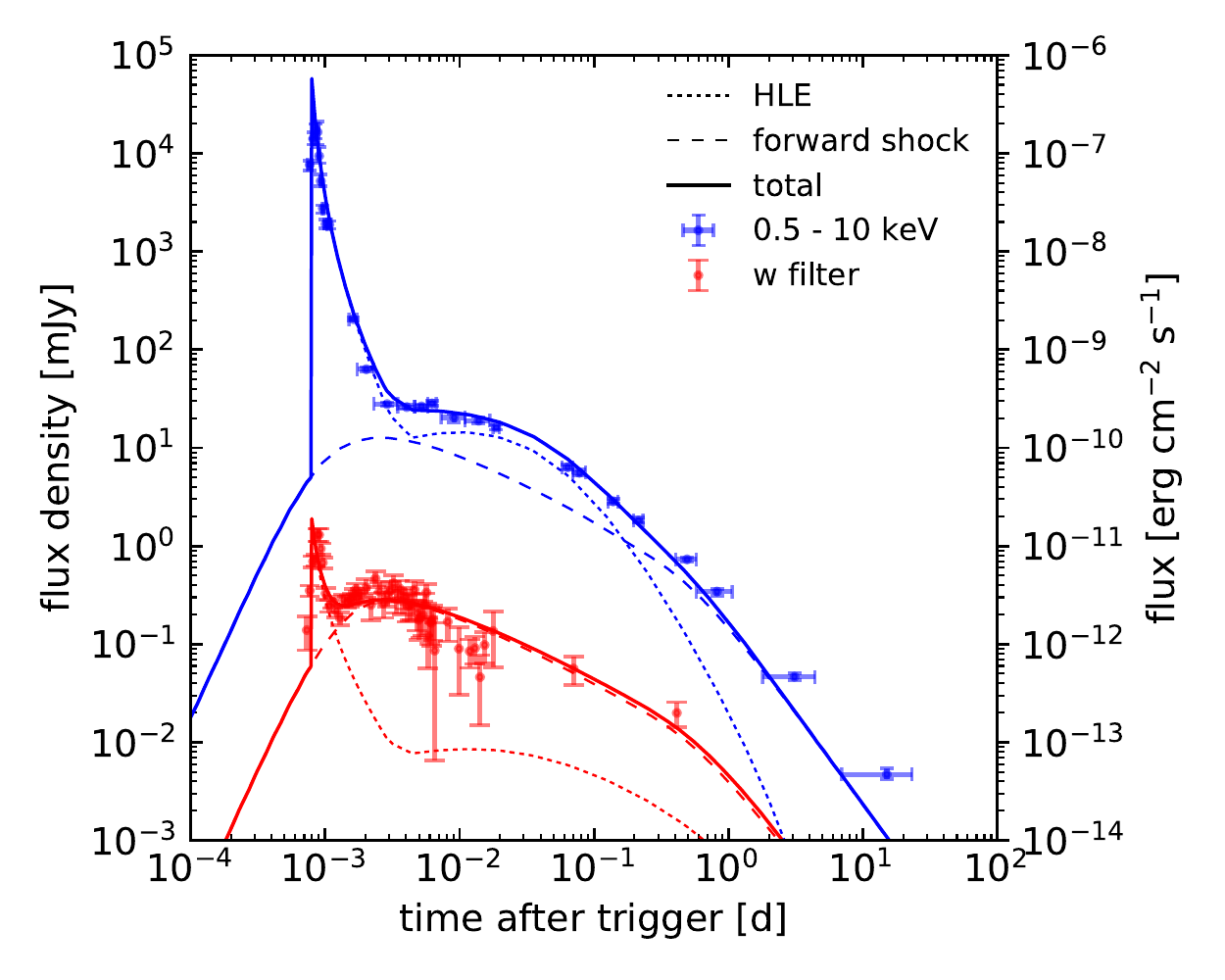}
    \caption{GRB 061121 lightcurves as observed by XRT (energy flux in the 0.5 -- 10 keV band, blue error bars) and UVOT (mean flux density in the white filter, red error bars, de-reddened assuming $E(B-V)=0.055$) compared to the predictions of our model (solid lines), where both the high latitude emission (dotted lines) and forward shock (dashed lines) emission are taken into account. The adopted parameters are reported in the text. The w filter model light curve accounts for the transmission curve of the filter, retrieved from the HEASARC website (\url{https://heasarc.gsfc.nasa.gov/docs/heasarc/caldb/swift/docs/uvot/})}.
    \label{fig:061121_modelling}
\end{figure}

In order to model the multiwavelength emission of the GRB 061121, we collected the optical data in the {\it Swift}/UVOT white filter from \cite{Page2007}. 
The observed magnitudes are corrected for the Galactic extinction \citep{Schlafly2011} and for extinction in the host galaxy.
We use a model including the high latitude emission and the forward shock afterglow from the same jet structure. The model with the X-ray and optical observations are shown in Fig.~\ref{fig:061121_modelling}. We assume a power-law jet structure with
$k_E=2$, $k_\Gamma=2.2$, $\Gamma_\mathrm{c}=180$, $\theta_\mathrm{c}=\theta_\Gamma=2^\circ$, $E_\mathrm{c}=3\times 10^{53}\,\mathrm{erg}$, $n=0.3\,\mathrm{cm^{-3}}$, $p=2.1$, $\epsilon_\mathrm{e}=0.1$, and $\epsilon_\mathrm{B}=10^{-4}$. Here $k_E$ and $k_\Gamma$ refer to the power law indices of the energy and Lorentz factor structures, respectively. As described in the previous section, we assume a 2BPL shape for the prompt emission intrinsic spectrum. We fix the comoving peak photon energy at $E_\mathrm{peak}'= 500\,\mathrm{keV}/\Gamma_\mathrm{c} = 2.8\,\mathrm{keV}$, the comoving low-energy break photon energy at $E_\mathrm{break}'= 3\,\mathrm{keV}/\Gamma_\mathrm{c} = 1.7\times 10^{-2}\,\mathrm{keV}$, the photon indices to $\alpha_1=-2/3$, $\alpha_2=-1/2$ and the high-energy photon index to $\beta=-4.1$. Finally, we assume a prompt emission radius $R_0=5\times 10^{15}\mathrm{cm}$, and we assume the high latitude emission to start $60\,\mathrm{s}$ after the GRB trigger time (i.e.~around the time of the main pulse). The resulting high-latitude emission in the X-rays dominates over the forward shock emission during the plateau. It is necessary to produce the initial steep decay in the X-ray and optical band, and the flatness and the duration of the X-ray plateau. On the other hand, the predicted high latitude  emission results to be negligible with respect to the forward shock in the optical band. The high latitude optical emission is faint because it corresponds to the low energy tail of the prompt emission spectrum. The high latitude emission and the forward shock expected from the structured jet are able to explain the chromatic behaviour of the optical and X-ray light curves. We test also a Gaussian profile for the jet structure finding  almost identical results.

\subsubsection{GRB 100906A}

\begin{figure}
    \centering
    \includegraphics[width=\columnwidth]{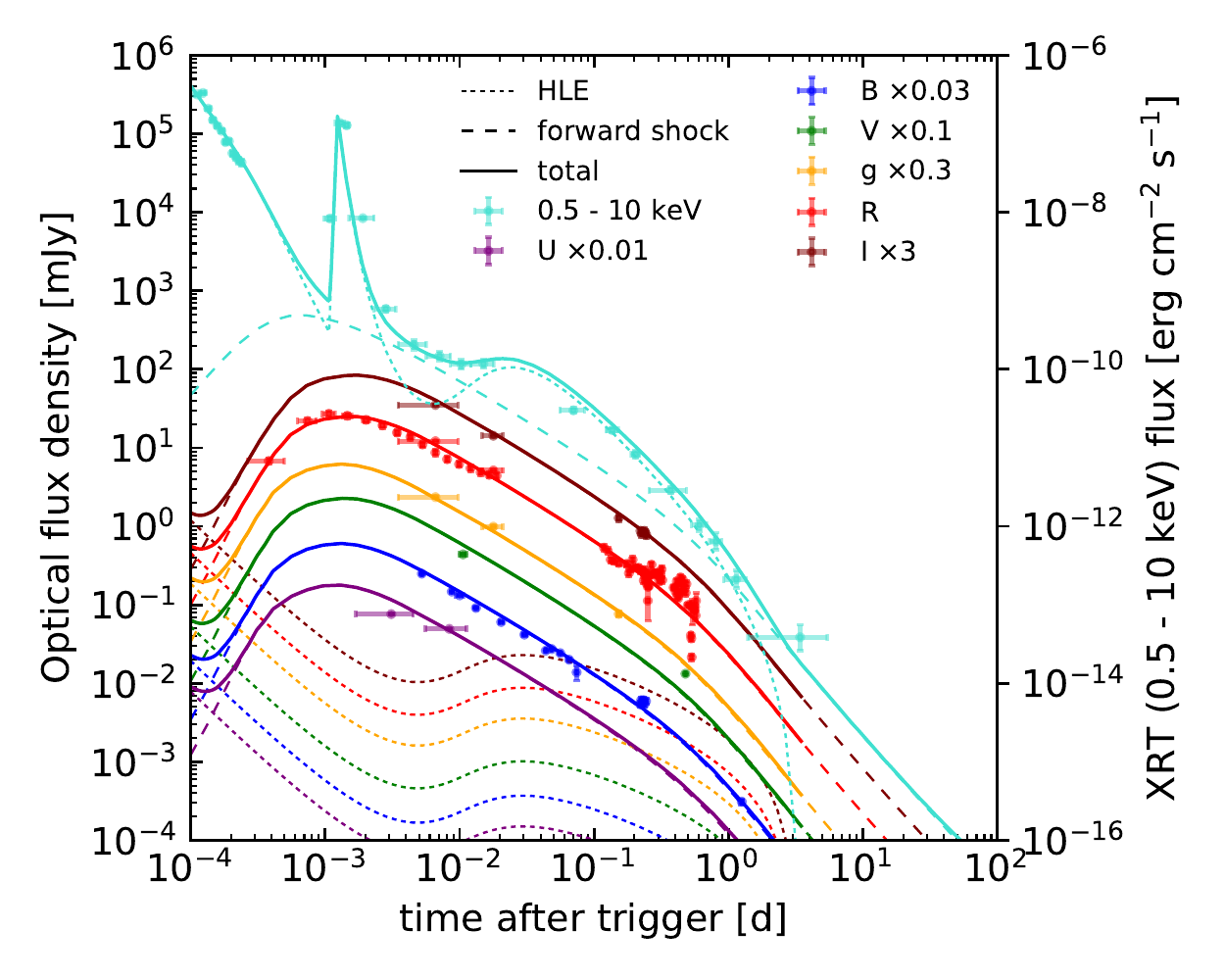}
    \caption{Same as Fig.~\ref{fig:061121_modelling}, but for GRB 100906A light curves as observed by XRT (energy flux in the 0.5 -- 10 keV band, cyan error bars) and in optical as observed by several facilities (in-band flux density, coloured error bars, rescaled for for better visualization by the factors shown in the legend). The adopted parameters are reported in the text.}
    \label{fig:100906A_modelling}
\end{figure}

We collected multi-filter optical data from \cite{Gorbovskoy2012} and private communication with A.~Melandri. The observed magnitudes are corrected for galactic extinction \citep{Schlafly2011} and for extinction in the host galaxy. Fig.~\ref{fig:100906A_modelling} shows the comparison between the observed light curves and our high latitude and forward shock emission model, obtained assuming a Gaussian jet structure with $\Gamma_\mathrm{c}=160$, $\theta_\mathrm{c}=4.2^\circ$, $\theta_\Gamma=3.3^\circ$, $E_\mathrm{c}=5\times 10^{53}\,\mathrm{erg}$, $n=15\,\mathrm{cm^{-3}}$, $p=2.1$, $\epsilon_\mathrm{e}=0.03$, and $\epsilon_\mathrm{B}=2\times 10^{-2}$. For the prompt emission, we assume again a 2BPL spectrum as before, with parameters taken from spectral fitting of the brightest BAT pulse. Taking the fitting parameters from \citet{gor2017}, we obtain $E_\mathrm{peak}'= 100\,\mathrm{keV}/\Gamma_\mathrm{c} = 0.625\,\mathrm{keV}$, $E_\mathrm{break}'= 10\,\mathrm{keV}/\Gamma_\mathrm{c} = 6.25\times 10^{-2}\,\mathrm{keV}$, and  $\alpha_1=-2/3$, $\alpha_2=-1/2$ as before. The high-energy photon index is set to $\beta=-3.2$ and we assume a prompt emission radius $R_0=3\times 10^{15}\mathrm{cm}$. A second, later emission event with an almost comparable peak flux is present in the XRT lightcurve at around $100\,\mathrm{s}$ post-trigger. The spectrum of this pulse can be fitted by the 2SBPL with $E_\mathrm{peak}\sim 5\,\mathrm{keV}$ and $E_\mathrm{break}\lesssim 1\,\mathrm{keV}$, with a high-energy spectral slope $\beta\sim -3.7$. Although it is subdominant, given the softer spectrum, we included the high latitude emission from this pulse for completeness. For this pulse, we adopt the same jet structure of the first one and we assume a starting time $t_0=85\,\mathrm{s}$ in the observer frame.  Also in this case, the combination of the high-latitude and forward shock emission is able to explain the features of the multiwavelength light curves (see Fig.~\ref{fig:100906A_modelling}), with the high-latitude emission dominating over the forward shock emission during the plateau phase in the X-rays, while being negligible in the optical range. Note that we find equally satisfying results with a power law structure, setting $\theta_\mathrm{c}=\theta_\mathrm{\Gamma}=2^\circ$ and using slopes $k_E=2$ and $k_\Gamma=3.1$. In this latter case, the structure would be very similar to the one we find for GRB 061121, the differences being almost entirely in the distance and ISM density, pointing to the quasi-universal jet structure hypothesis \citep{Salafia2015,Salafia2019b,Salafia2019}.

\subsection{Spectral properties of the high-latitude emission plateau}
Predictions of the high-latitude emission model extend to the spectral properties of GRB X-ray counterparts. 
Since in our model the high latitude emission gives an essential contribution to the X-ray emission in the plateau phase, we expect that the GRB X-ray spectrum includes properties of the prompt emission spectrum. Assuming that the comoving spectrum is the same at all angles $\theta$, the decrease of $\mathcal{D}$ with time means that the observer is probing a progressively higher energy part of the prompt emission spectrum, since the observed photon energy is $E_{\rm obs} = E'\mathcal{D}$. 
For the Gaussian jet structure with $\Gamma_{c} \sim$ 100, our results indicate that $\mathcal{D} \sim \rm 10$ at the start of the plateau. Therefore, the spectral index in the plateau phase (at $\sim$ 10 keV) would correspond to that at around $\sim$ 100 of keV in the prompt emission pulse at $t_0$. This is close to the typical peak energy of the prompt emission spectrum ($\sim 200$ keV; e.g. \citealt{Nava2011}). The photon index in the plateau phase can vary between $-1.5$, the spectral slope below $E_{\rm peak}$, and $\hat{\beta_{p}}$ which is the photon index above $E_{\rm peak}$. The value of $\hat{\beta_{p}}$ ranges between -2 and -3 (e.g. \citealt{Nava2011}) which is consistent with the spectral analysis of the plateau phase found by \citealt{Liang2007}, who suggest a range of photon indices between -1.5 and -2.5.

\section{Conclusions}
We studied the high latitude emission arising from a switched off pulse of relativistic jets with an angular structure in the bulk motion and comoving brightness. Using both Gaussian and power-law jet structures, we tested the predictions of our model by comparing them with the X-ray and optical light curves of the GRB 061121 and GRB 100906A. Our results are summarized in the following:
\begin{itemize}
    \item The plateau phase observed in a good fraction of X-ray GRB lightcurves can arise from the high-latitude emission of a structured jet. Plateaus starting at $t_{\rm obs} \sim 10^{2}-10^{3}$ s require that the size of the jet at the start of the plateau be $R_{0} \gtrsim 10^{15}$ cm. While in the Gaussian jet structure only plateau durations of $\sim 10^{3}$ s can be obtained, the power-law jet structure can provide more extended plateaus (up to $\sim$ few $\times 10^{4}$ s). However, changing the emission region and jet parameters the model can account for longer and brighter X-ray plateaus as observed by {\it Swift}/XRT. 
    \item The high latitude emission model from a structured jet is expected to produce two further segments during the flux decay in the post-plateau phase: a power-law followed by a very sharp drop. The sharp drop can provide a novel explanation for such puzzling feature observed in some GRB X-ray lightcurves (e.g. \citealt{Troja2007}). 
    \item Adding the high latitude emission to the radiation from the forward shock enables to account for the chromatic behaviour of the lightcurves in the optical and X-ray bands. This is mainly due to the interplay between two separated emission regions: the X-ray plateau comes from the prompt emitting wings of the jet while the optical is most likely dominated by forward shock emission.
    \item The spectra of the plateau phase in the X-ray energy range are consistent with the high-latitude emission predictions. In particular, measured photon indices ($-2.5<\alpha<-1.5$, \citealt{Liang2007}) are in agreement with the synchrotron model for the prompt emission. The spectral softening typically observed in the steep-decay/plateau transition is also naturally explained in our model.
\end{itemize}
Previous studies of high-latitude emission from inhomogenous jets \citep{Dyks2005,Yamazaki2006,Takami2007} were focused on the influence of jet structure on the fast decay phase. The flattening of high-latitude emission at $\sim \rm 30$ s due to the bulk motion structure has been noticed previously in \citet{Dyks2005}. However, these authors did not obtain results for the plateau phase observed at $\sim 10^{2}-10^{3}$ s. This is due to the limited parameter space considered in their work, e.g.~the size of the jet, fixed to $R_{0} \sim 10^{14}$ cm.  

Here, we have shown that the high-latitude emission from a structured jet is able to explain the complex morphology of X-ray counterparts of GRBs, including fast decay, plateau and post-plateau phases. The high latitude emission added to radiation from the forward shock is able to produce the shape, the luminosity and the duration of the X-ray emission and to account for the chromatic behaviour of the X-ray/optical lightcurves.
The systematic application of this model to optical-to-X-ray data 
will provide a mean of probing the angular structure of GRB jets and the size of the emitting region during the prompt emission. A thorough exploration of these aspects will be the subject of subsequent works.

\acknowledgements
We would like to thank Andrea Melandri for sharing the optical data. GO is thankful to Gabriele Ghisellini and Elias S. Kammoun for fruitful discussions. MB, SDO, GO acknowledge financial contribution from the agreement ASI-INAF n.2017-14-H.0.  GO and SA are thankful to INAF -- Osservatorio Astronomico di Brera for kind hospitality during the completion of this work. SA acknowledges
the GRAvitational Wave Inaf TeAm - GRAWITA (P.I. E. Brocato) and the PRIN-INAF "Towards the SKA and CTA era: discovery, localization and physics of transient sources". This work made use of data supplied by the UK Swift Science Data Centre at the University of Leicester.

\appendix
\section{Light curve of the high latitude emission from a structured outflow \label{appendix}}

The observed specific flux of the source is defined in the following way: 

\begin{equation}
F_{\nu} = \int d\Omega_{\rm obs} I_{\nu} (\nu, t) \cos(\theta_{\rm obs})
\label{eq:flux_general}
\end{equation}

where $\Omega_{\rm obs}$, $I_{\nu}$ and $\theta_{\rm obs}$
are the observed solid angle, the observed specific intensity and the angle between the normal of the observing surface of the instrument and the element of the surface from where the photons come.

We further assume the object of our interest, i.e. an instantaneous emission produced throughout a  structured outflow at a given time. Due to the infinitesimally short duration of the pulse, at a given time the observer receives the emission from a certain equal arrival time ring from the surface of the outflow. Therefore, given that the specific intensity in the lab frame trasforms like $I_\nu(\nu, t) = D^3(\theta) I'_{\nu'}(\nu', t)$, the general eq. \ref{eq:flux_general} for an observed aligned with the center of the outflow returns 

\begin{equation}
F_{\nu}(t_{\rm obs}) = \frac{2\pi}{d_{L}^{2}} \int^{\pi/2}_0 
\eta'_{\nu'}(\nu') \delta(t-t_{\rm em}) D^{3}(\theta) R^{2}(\theta) \sin\theta \cos\theta d\theta
\label{eq:flux_special}
\end{equation}

where $d_{L}$ is the luminosity distance to the source, $R(\theta)$ is the radius of the outflow measured from its origin. We wrote the specific intensity in the comoving frame as $I'_{\nu'}(\nu', t) = \eta'_{\nu'}(\nu')\delta(t-t_{em})$, where $\eta'_{\nu'}$ is the energy emitted per unit area, per unit frequency, per unit solid angle, at an angle $\theta$, as measured in the comoving frame, $\delta(t-t_{em})$ is the Dirac delta function.  At a given time $t_{em}$ measured in the lab frame, the infinitesimally short-duration pulse is produced throughout the entire outflow with the same spectral shape $S(\nu')$. At an observer time $t_{obs}= t[1-\beta(\theta)\cos \theta ] =t/D(\theta)\Gamma(\theta)$, the specific flux is given by :

\begin{equation}
    F_\nu (t_{\rm obs}) = \frac{2\pi}{d^2_L}\frac{1}{t_{\rm obs}}\int^{\pi/2}_0 \eta'_{\nu'}(\nu')D^3(\theta)\frac{\delta(\theta - \theta_{\rm obs})}{D^2(\theta_{\rm obs})\Gamma^2(\theta_{\rm obs})\left|\displaystyle \frac{d\beta}{d\theta}\cos\theta - \beta\sin\theta\right|_{\theta = \theta(t_{\rm obs})}}R^2(\theta)\sin \theta \cos \theta d\theta
    \label{eq:step1}
\end{equation}

where we applied the following transformation for the Dirac delta (which corresponds to the standard transformation rule for the Dirac delta of a function):
\begin{equation}
    \delta(t-t_{\rm em}) = \delta[t_{\rm obs}D(\theta)\Gamma(\theta) - t_{\rm em}] = \frac{\delta(\theta -\theta_{\rm obs})}{\left|\displaystyle \frac{d}{d\theta}\left[t_{\rm obs}D(\theta)\Gamma(\theta) - t_{\rm em}\right]\right|_{\theta = \theta(t_{\rm obs})}}
\end{equation}
$\theta(t_{obs})$ is the angle where the photons reaching the observer at a given $t_{obs}$ are emitted from (which corresponds to the solution of $t_{obs}D(\theta)\Gamma(\theta) = t_{em}$). The derivative at the denominator can be written explicitly as:
\begin{equation}
    \frac{d}{d\theta} \left[t_{\rm obs}D(\theta)\Gamma(\theta) - t_{\rm em}\right]= 
    t_{\rm obs}\frac{d}{d\theta} [D(\theta)\Gamma(\theta)] = t_{\rm obs}D^2\Gamma^2\Bigl(\frac{d\beta}{d\theta}\cos\theta - \beta\sin\theta\Bigr)
\end{equation}

With the aid of Dirac delta the solution of the integral in Eq. \ref{eq:step1} is straightforward and writes as:
\begin{equation}
    F_\nu(t_{\rm obs}) = \frac{2\pi}{d^2_L}\frac{R^2_0}{\beta^2_0}\frac{D^2(\theta(t_{\rm obs}))}{\Gamma(\theta(t_{\rm obs}))}\eta'_{\nu'}(\nu') \frac{\beta^2(\theta(t_{\rm obs}))\sin\theta(t_{\rm obs})\cos\theta(t_{\rm obs})}{t_{\rm obs}D(\theta(t_{\rm obs}))\Gamma(\theta(t_{\rm obs}))\left|\displaystyle \frac{d\beta}{d\theta}\cos\theta - \beta\sin\theta\right|_{\theta = \theta(t_{\rm obs})}}
    \label{eq:step2}
\end{equation}
where we used $R(\theta(t_{\rm obs})) = R_0\beta(\theta(t_{\rm obs}))/\beta_0$.

We can further simplify Eq. \ref{eq:step2} by noticing that:
\begin{equation}
    \frac{d\theta}{dt_{\rm obs}} = - \frac{1}{t_{\rm em}\left(\displaystyle \frac{d\beta}{d\theta}\cos\theta - \beta\sin\theta\right)} = \frac{1}{t_{\rm obs}D(\theta)\Gamma(\theta)\left|\displaystyle \frac{d\beta}{d\theta}\cos\theta - \beta\sin\theta)\right|}
\end{equation}
This equation is obtained from $t_{\rm obs}D(\theta)\Gamma(\theta) = t_{em}$. Considering that $\beta(\theta)$ is decreasing with $\theta$, the argument of the module at the denominator is negative defined.

Substituting this last equation in \ref{eq:step2} we obtain:

\begin{equation}
F_{\nu}(t_\mathrm{obs}) =  \frac{2\pi}{d_{L}^{2}} \,
\frac{R_{0}^{2}}{\beta^{2}_0} \,\Bigl(
\mathcal{D}^{2}(\theta) \, S(\nu') \, \epsilon(\theta)  \, \frac{\beta^{2}(\theta)}{\Gamma(\theta)} \, \cos\theta\sin\theta\frac{ d\theta}{dt_{\rm obs}} \, \Bigr)_{\theta = \theta(t_{\rm obs})}
 \label{eq:flux_final}
\end{equation}

We have introduced the angular dependence of the comoving brightness, i.e. we set $\eta'_{\nu'} = \epsilon(\theta) S(\nu')$. We also used the eq. \ref{eq:time} to simplify final expression and to represent it in terms of the differential area $\sin\theta d\theta /dt_{\rm obs}$. 
The observed time $t_{\rm obs}$ is measured from the imaginary photon emitted at $R=0$. However, one can easily change the reference time to the time at $R_{0}$ by introducing a delay of $R_{0}/c \left[1/\beta(0)-1\right]$. It is worth noticing that the delta function that appears in $I'_{\nu'}$ is defined in the lab frame. This is the proper frame to define an instantaneous emission. It can be expressed as a delta function in the comoving frame  $\delta(t-t_{\rm em})= \Gamma(\theta)\delta(t'-t'_{\rm em})$, where $t'-t'_{\rm em} = \Gamma(\theta)(t-t_{\rm em})$ is the time interval in the comoving frame. By repeating all the previous calculations with this expression leads to the same expressions of Eq. $\ref{eq:flux_final}$.

In order to make as most explicit as possible the dependence of Eq. \ref{eq:flux_final}
on $t_{\rm obs}$. To this aim noticing that $D(\theta(t_{\rm obs})) = t_{\rm em}/(t_{\rm obs}\Gamma(\theta(t_{\rm obs})))$ and $t_{\rm em} = R_0/(\beta_0 c)$ we can write:
\begin{equation}
    F_{\nu}(t_\mathrm{obs}) =  \frac{2\pi}{d_{L}^{2}} \,
\frac{R_{0}^{4}}{\beta^{4}_0c^2}\Bigl(S(\nu') \, \epsilon(\theta)  \, \frac{\beta^{2}(\theta)}{\Gamma^3(\theta)} \, \cos\theta\sin\theta\frac{ d\theta}{dt_{\rm obs}} \, \Bigr)_{\theta = \theta(t_{\rm obs})}t^{-2}_{\rm obs}
\label{eq:finalflux2}
\end{equation}
If we consider a power-low spectrum $S_{\nu'} = S_{\nu_0^{'}}\nu'^{\beta_s}_0\nu'^{-\beta_s}$, where $\beta_s$ is the spectral index, the general equation Eq. \ref{eq:finalflux2} becomes:
\begin{equation}
     F_{\nu}(t_\mathrm{obs}) =  \frac{2\pi}{d_{L}^{2}}\frac{R^2_0}{\beta^2_0}
\Bigl(\frac{R_0}{\beta_0c}\Bigr)^{2+\beta_s}S_{\nu'_0}\nu'^{\beta_s}_0\Bigl(\epsilon(\theta)  \, \frac{\beta^{2}(\theta)}{\Gamma^{3+\beta_s}(\theta)} \, \cos\theta\sin\theta\frac{ d\theta}{dt_{\rm obs}} \, \Bigr)_{\theta = \theta(t_{\rm obs})}\nu^{-\beta_s}t^{-(2+\beta_s)}_{\rm obs}
\label{eq:finalflux3}
\end{equation}
This last equation shows the power-law temporal behavior of \citet{Kumar2000} (see also \citet{Uhm2015} and \citet{Kumar2015}) but there is a further time dependent factor which comes from the structure of the jet and it is dependent on $\theta$ in Eq. \ref{eq:finalflux3}. The solution of \citet{Kumar2000} is thus recovered in the spherical case (setting $\epsilon(\theta) = 1$ and $\cos \theta(t_{obs})\simeq 1$):
\begin{equation}
        F_{\nu}(t_\mathrm{obs}) =  \frac{2\pi}{d_{L}^{2}}R_0
\Bigl(\frac{R_0}{\beta_0c}\Bigr)^{2+\beta_s}S_{\nu'_0}\nu'^{\beta_s}_0  \, \frac{c}{\Gamma^{3+\beta_s}}  \, \nu^{-\beta_s}t^{-(2+\beta_s)}_{obs}
\end{equation}

\bibliographystyle{aasjournal} 
\bibliography{references} 
\end{document}